\documentclass[%
 reprint,
superscriptaddress,
 amsmath,amssymb,
 aps,
prapplied,
]{revtex4-2}

\usepackage{graphicx}
\usepackage{dcolumn}
\usepackage{bm}
\usepackage{enumitem}
\usepackage{xcolor}
\usepackage{todonotes}
\usepackage{textgreek}
\usepackage{booktabs}
\usepackage{multirow}
\newcommand{\ff}{f\mkern-4mu f}

\makeatletter\let\selectlanguage\@gobble\makeatother

\graphicspath{{figures/}}

\begin{document}

\preprint{APS/123-QED}

\title{Large-scale characterisation of silicon T centres with tuning-free broadband cavities}

\author{Christian~F.~Primavera}
\affiliation{QuTech and Kavli Institute of Nanoscience Delft, Delft University of Technology, 2628 CJ Delft, The Netherlands}

\author{Matteo~Pirro}
\affiliation{QuTech and Kavli Institute of Nanoscience Delft, Delft University of Technology, 2628 CJ Delft, The Netherlands}
\affiliation{Department of Quantum and Computer Engineering, Delft University of Technology, 2628 CJ Delft, The Netherlands}

\author{Elena~Volkova}
\affiliation{QuTech and Kavli Institute of Nanoscience Delft, Delft University of Technology, 2628 CJ Delft, The Netherlands}
\affiliation{Department of Quantum and Computer Engineering, Delft University of Technology, 2628 CJ Delft, The Netherlands}

\author{Joan~Mascaró~Rosselló}
\affiliation{QuTech and Kavli Institute of Nanoscience Delft, Delft University of Technology, 2628 CJ Delft, The Netherlands}

\author{Cas~C.P.E.~Hoppenbrouwers}
\affiliation{QuTech and Kavli Institute of Nanoscience Delft, Delft University of Technology, 2628 CJ Delft, The Netherlands}

\author{Bram~Zijlstra}
\affiliation{QuTech and Kavli Institute of Nanoscience Delft, Delft University of Technology, 2628 CJ Delft, The Netherlands}

\author{Ronald~Hanson}
\affiliation{QuTech and Kavli Institute of Nanoscience Delft, Delft University of Technology, 2628 CJ Delft, The Netherlands}

\author{Carlos~Errando-Herranz}
\email{C.ErrandoHerranz@tudelft.nl}
\affiliation{QuTech and Kavli Institute of Nanoscience Delft, Delft University of Technology, 2628 CJ Delft, The Netherlands}
\affiliation{Department of Quantum and Computer Engineering, Delft University of Technology, 2628 CJ Delft, The Netherlands}

\date{\today}

\begin{abstract}
T centres in silicon combine an optical transition in the telecom O-band with a long-lived electron and nuclear spin in a photonics- and CMOS-compatible host. However, weak emission of photons in the zero-phonon line (ZPL) makes measurements slow. Cavities can increase the ZPL photon emission, but high-Q cavities require individual tuning to the T centre wavelength, hindering statistical understanding of T centre properties. Here, we overcome this limitation by developing low-Q confocally-addressable cavities. We create an array of 400 cavities, detuned $0.0\pm1.1$~nm from the T centre, with an average quality factor of $298\pm78$. The combination of high accuracy and moderate quality factors removes the need for individual tuning. The array yields a median T centre signal increase of $6.5\times$, corresponding to a $42\times$ measurement speedup. Studying the cavity statistics enables new measurements: comparing to a Monte Carlo simulation we infer a T centre density in the sample of $4.2^{+1.3}_{-0.8}$~\textmu m$^{-2}$, without relying on an estimate for the setup collection efficiency.
\end{abstract}


\maketitle

\section{Introduction}
Building a quantum network requires coupling stable stationary qubits to flying qubits that can be transmitted over distance with high fidelity \cite{divincenzo_physical_2000,wehner_quantum_2018}. This can be achieved by coupling long-lived spins in the solid state to a photon \cite{pompili_realization_2021,awschalom_quantum_2018,ruf_quantum_2021,wolfowicz_quantum_2021}. The T centre in silicon is an attractive candidate for a quantum network node for multiple reasons. T centres can be created in silicon-on-insulator (SOI), an established commercial photonics platform, which creates a viable route to on-chip integration \cite{soref_past_2006,simmons_scalable_2024,sandholzer_single-photon_2026,macquarrie_generating_2021,jhuria_programmable_2024,andrini_activation_2024,zhiyenbayev_scalable_2023}. Silicon can also be engineered to a high degree of isotopic purity, creating a clean magnetic environment \cite{higginbottom_memory_2023}. Moreover, T centres couple directly to the telecom O-band, and have long electron and nuclear spin coherence times \cite{dhaliah_first-principles_2022,bergeron_silicon-integrated_2020,higginbottom_optical_2022,higginbottom_memory_2023,brunelle_silicon_2026,song_long-lived_2025,song_entanglement_2026}.

The performance of a quantum network scales with the rate of photons extracted from the zero-phonon line (ZPL), the kind that can be used for generating entanglement. In the case of the T centre, achieving a high ZPL count rate is challenging because it intrinsically has a low quantum efficiency (QE) (around 0.2 \cite{clear_optical-transition_2024}) and Debye-Waller factor (DWF) (around 0.23 \cite{bergeron_silicon-integrated_2020}), combined with a long radiative lifetime of 0.94~\textmu s \cite{bergeron_silicon-integrated_2020}. Furthermore, silicon has a high refractive index.

Performance has been significantly increased using waveguides \cite{deabreu_waveguide-integrated_2023,lee_high-efficiency_2023} or cavities with high quality factor (Q) \cite{quan_deterministic_2011,islam_cavity-enhanced_2024,johnston_cavity-coupled_2024,saggio_cavity-enhanced_2024,lefaucher_bright_2025}, but both run into limits for scalable characterisation. Waveguides have no significant effect on the QE and DWF \cite{deabreu_waveguide-integrated_2023}. High-Q cavities perform well in this regard, but have a linewidth that is much narrower than the fabrication precision, creating the need for individual tuning \cite{johnston_cavity-coupled_2024}. To scale up T centre characterisation, fast confocal imaging is necessary \cite{sutula_large-scale_2023}, but weak ZPL emission hampers this.

Here, we overcome this challenge by developing low-Q, confocally-addressable and tuning-free cavities aligned to the T centre by design. This allows us to measure photoluminescence of 400 T centre sites with 42 times shorter integration time than in unpatterned samples, shifting the measurement time bottleneck to the automated setup alignment. We use the resulting statistics to estimate the T centre density without relying on estimates of the collection efficiency of the setup.

\section{Broadband cavity design}
\label{sec:design}
\begin{figure*}
    \centering
    \includegraphics[width=\linewidth]{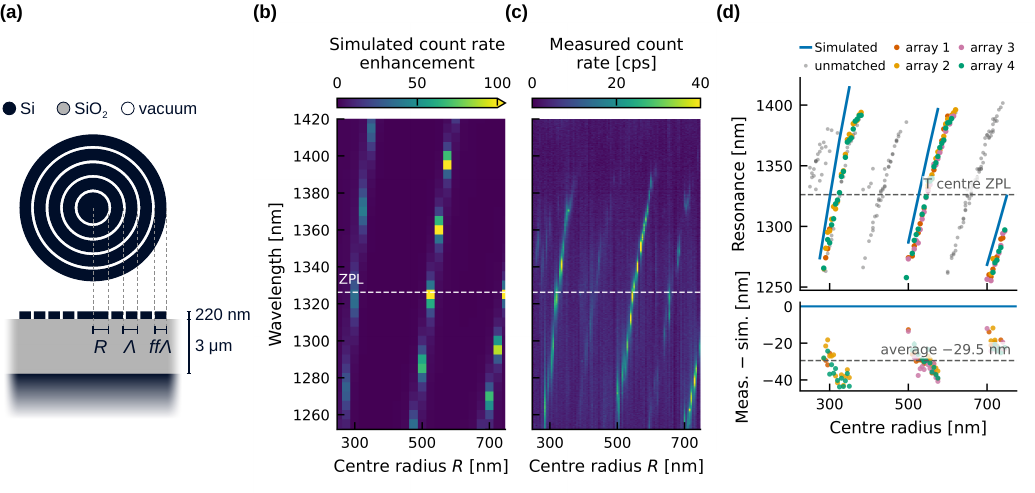}
    \caption{\label{fig:design}%
    \textbf{Cavity design, simulation, and experimental verification.} (a) Cavity geometry top view: central disc radius $R$, grating period $\Lambda$, fill factor $\ff$. The side view below shows the 220~nm SOI device layer on a 3~\textmu m layer of oxide (not to scale). (b) Simulated cavity figure of merit: count rate enhancement ($E = F_P\ \eta_c/\eta_0$) versus emission wavelength and radius $R$, with $\Lambda=520$~nm, $\ff=0.8$. The dashed line indicates the T-centre ZPL. The cavity modes shift with $R$. (c) High-power (3~mW) PL measurement on 100 cavities in one of the four arrays. Under high-power excitation, short-lived broadband background fluorescence from the material drowns out the saturated T centre signal. The measured signal is the broadband fluorescence modulated by the cavity count rate enhancement. (d) Fitted resonance of every detected mode in all four arrays (colours) versus the simulation (lines). Lower panel: difference between matched modes in measurement and simulation. More details in Appendix Fig.~\ref{fig:app-sweep-maps} and \ref{fig:app-sweep-stats}.}
\end{figure*}

For enhancing the T centre, we use `bullseye' cavities, also known as circular Bragg gratings \cite{rickert_optimized_2019,rickert_high_2025}. A bullseye cavity consists of a centre disc, surrounded by concentric rings [Fig.~\ref{fig:design}(a)]. It achieves the increase in collected light by a combination of Purcell enhancement and collection efficiency improvement. The periodic nature of the rings creates a weak mirror, which confines the mode to the disc and creates some Purcell enhancement. Simultaneously, the rings diffract the radiated light out of the plane, sending far more emission into the collecting objective than what would radiate out of the slab mode of unpatterned SOI.

In our design, the bullseye cavity is defined by centre radius $R$, grating period $\Lambda$, and filling factor $\ff$. The device layer thickness of 220~nm is fixed by the unpatterned wafer we start from. These parameters will be optimised to maximise the collected ZPL light, which is proportional to the product of the Purcell factor $F_P$ and the collection efficiency $\eta_c$ into an objective with NA=0.85.

The number of rings is also of importance, since adding rings will lead to a better confinement of the light. This is relatively easy in fabrication, but computationally intensive in the simulations. We therefore fabricate $N=5$ rings, but simulate only at $N=3$ (see Appendix Fig.~\ref{fig:app-simulation}(b) for more detail).

For our system, we deliberately choose a low quality factor ($Q$). Purcell enhancement grows linearly with $Q$, but only if the mode overlaps with the emitter wavelength. Assuming the emitter resonance to be a delta peak at wavelength $\lambda_e$, which is justified in this low-Q regime, the wavelength overlap $L$ with a Lorentzian cavity mode with linewidth $\Gamma$, detuned by $\Delta \lambda$, is given by $L=(\Gamma/2)^2/[\Delta\lambda^2+(\Gamma/2)^2]$. The enhancement scales with $QL$, so we should optimise for
\begin{equation}
  QL = \frac{\lambda_\mathrm{e}}{\Gamma}
         \frac{(\Gamma/2)^2}{\Delta\lambda^2+(\Gamma/2)^2}.
  \label{eq:figure-of-merit}
\end{equation}
If we maximise over $\Gamma$ for a fixed detuning, we find $\Gamma=2|\Delta\lambda|$: the linewidth at FWHM should be twice the detuning. For good cavity performance, but also a high yield of resonant cavities, the sweet spot would be to design for $\Gamma \approx 2\sigma_{\Delta\lambda}$, with $\sigma_{\Delta\lambda}$ the fabrication scatter of cavity mode wavelength detuning.

To design bullseye cavities that are resonant with T centres, we simulate their properties in a COMSOL EMW frequency domain study. The figure of merit is the count rate enhancement $E$, which is the product of the Purcell factor $F_P$ and the collection efficiency improvement $\eta_c/\eta_0$.

The filling factor $\ff$ is kept constant at 0.8 throughout the simulations, since it leads to Q factors in the desired range (see Appendix Fig.~\ref{fig:app-simulation} for more details). The period $\Lambda$ determines the range of wavelengths for which cavity modes are well-defined, see Appendix Fig.~\ref{fig:app-periods}. 

The radius $R$ can be used to tune the resonance wavelength of the cavity mode. It should be noted that varying any of the parameters will shift the resonance wavelength, but $R$ is the most direct tuning knob, as it influences other properties only slightly. The effect of changing the radius of the bullseye cavity is shown in Fig.~\ref{fig:design}(b), for $\Lambda=520$~nm, $\ff=0.8$.

The final bullseye cavity design was chosen from the middle branch of modes around $R=500-600$~nm, since this gives the maximum enhancement. For the branch at smaller radii, the quality factor is worse, and for larger radii, the mode volume $V$ becomes larger, also decreasing the enhancement. The calibration array was fabricated at $\Lambda=520$~nm, which is why that subset of the simulations is shown here. The final design parameters for the cavity are given in Table \ref{tab:final_cavity}.

\begin{table}[h]
\centering
\caption{Final simulated cavity parameters.}
\label{tab:final_cavity}
\begin{tabular}{llc}
\toprule
 & Parameter & Value \\
\midrule
\multirow{4}{*}{Settings} & $R$       & 535\,nm \\
                          & $\Lambda$ & 500\,nm \\
                          & $\ff$      & 0.8 \\
                          & $N$       & 5 \\
\midrule
\multirow{2}{*}{Simulation results}  & $Q$       & 221.6 \\
                          & $F_P$     & 22.6 \\
\bottomrule
\end{tabular}
\end{table}

Note that the result for the Purcell factor $F_P$ holds for a T centre in the middle of the cavity (at the mode maximum) with perfect polarisation overlap. The dependence on exact location is analysed in more depth in Appendix Fig.~\ref{fig:app-simulation}(a).

\section{Fabrication calibration}
\label{sec:sweep}

The exact design parameters are expected to encounter fabrication imperfections when being transferred onto the sample. We do an experimental calibration step for two reasons: to check whether the simulations are realistic qualitatively, and to quantitatively determine the correction we should apply in the design, so that the fabricated cavities are resonant with the T centre.

We fabricate four identical arrays of 100 cavities each, $\Lambda=520$~nm, $\ff=0.8$, with radius $R$ swept between 250-745~nm in steps of 5~nm. We characterise these using 3~mW of green laser excitation (515~nm), with collection of the photoluminescence (PL) in a spectrometer. The high excitation power causes short-lived broadband background fluorescence to dominate the signal. The measured spectrum is made up of the broadband fluorescence, modulated by the cavity count-rate enhancement profile. The measurement of one array is shown in Fig.~\ref{fig:design}(c), and the data for all arrays are shown in Appendix Fig.~\ref{fig:app-sweep-maps}.

Fig.~\ref{fig:design}(d) shows a comparison between the fitted cavity resonance wavelengths of the simulations and the four measured arrays. The simulations and measurements follow the same pattern, but with a clear offset of $-29.5 \pm 1.0$~nm. The offset is very similar between the arrays (see Appendix Fig.~\ref{fig:app-sweep-stats}), suggesting that it can be systematically compensated for. However, all four arrays are on the same sample, so variance between sample fabrication runs could still cause differences.

Apart from the modes that clearly match the simulated behaviour, there are extra resonance families in the measurements that did not show up in the simulations. These result from leaky whispering-gallery modes in the bullseye cavity rings (see Appendix \ref{app-simulation}).

From the initial fabrication sweep, the measured resonance crossing at the T centre emission wavelength allows us to extract a systematic radius offset of $20.2 \pm 0.8$~nm. Accounting for further optimization of the proximity effect correction (PEC), pattern density variations, and tool-related process changes, the total applied correction is $30$~nm, setting the final target design radius to $565$~nm.

To tackle residual fabrication uncertainty intrinsic to e-beam lithography and the nanofabrication process, an array of 9 devices was created, spanning radii in $2.5$~nm steps, centred on $565$~nm. This sweep was additionally replicated at a slightly different base dose and offset correction, to further compensate for process variability. The array at $R = 562.5$~nm was found to be resonant, deviating only $2.5$~nm from the corrected design target.

The resonance shifts by $\approx 1.14$~nm per nm of radius near the T centre resonance (see Fig.~\ref{fig:design}(d)), so each $2.5$~nm radius step moves the resonance by $2.9$~nm. Since this step is smaller than the average cavity linewidth $\Gamma = 4.7$ nm, the closest array always lies within $1.5$~nm of the ZPL, less than a third of the linewidth. The SEM imaging (see Appendix Fig.~\ref{fig:app-device}) confirms that the resonant array (designed at $R = 562.5$~nm) corresponds to a fabricated radius of $535$~nm and a filling factor of $0.8$, in full agreement with the ideal design parameters up to SEM resolution (See Table~\ref{tab:final_cavity}).

\begin{figure*}
    \centering
    \includegraphics[width=0.9\linewidth]{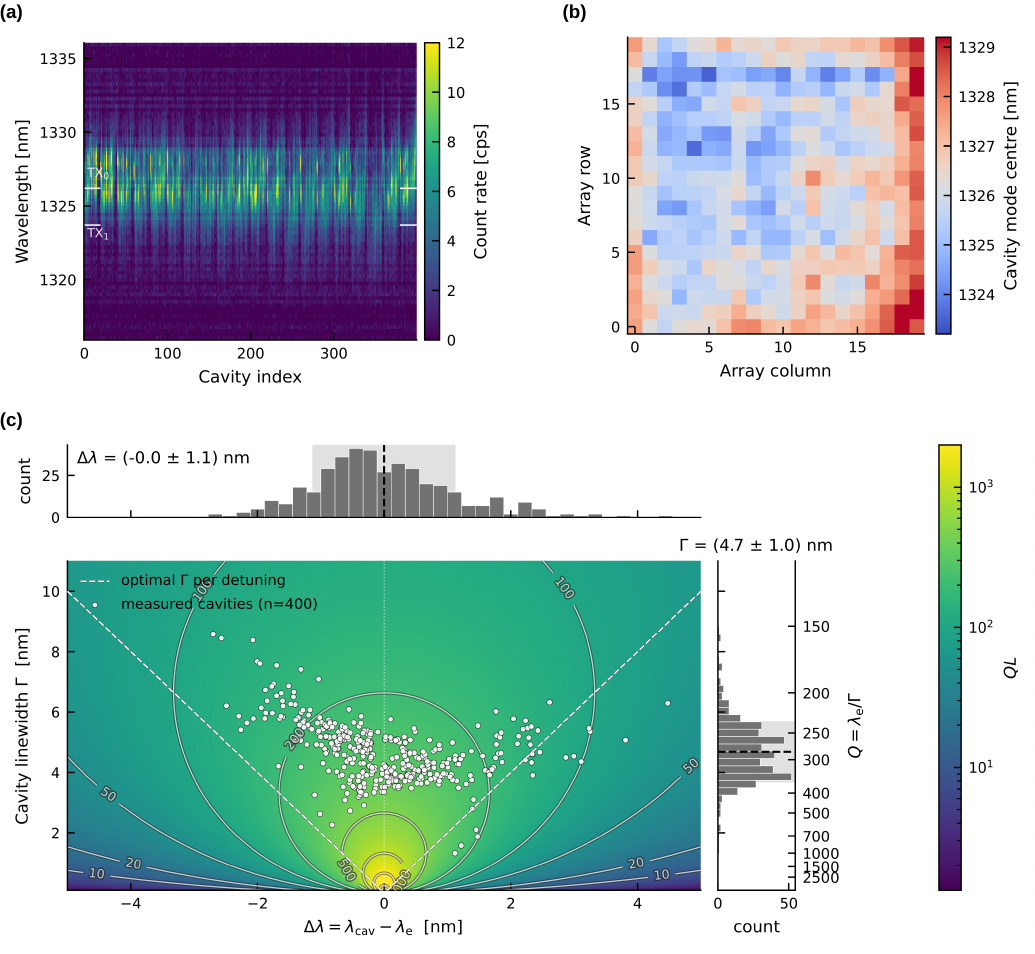}
    \caption{\label{fig:fab}%
    \textbf{Array of 400 identical cavities.} (a) PL spectra of all cavities ($\Lambda=500$~nm, $R=562.5$~nm, $\ff=0.8$). Wavelengths of the $TX_0$ and $TX_1$ T centre transitions are indicated. (b) Fitted cavity-mode centre on the physical array. This suggests there is a small residual proximity effect remaining from fabrication after our correction. (c) Spread of the 400 measured modes over detuning (x-axis) and linewidth at FWHM (y-axis). Colour scale shows $QL$, which is proportional to our performance metric. Contours = constant $QL$. Dashed lines indicate $\Gamma=2|\Delta\lambda|$, the linewidth that optimises $QL$ for a certain detuning. Distribution averages: $\Delta\lambda=(0.0\pm1.1)$~nm, $\Gamma=(4.7\pm1.0)$~nm, $Q=298\pm78$.}
\end{figure*}

\section{Large-scale fabrication of resonant cavities}
\label{sec:fab}

With the process calibrated we measure the $20\times20$ array of 400 identical cavities: $\Lambda=500$~nm, $\ff=0.8$, $R_{design}=562.5$~nm, $R_{real}=535$~nm. A high-power PL measurement shows the 400 cavity modes in Fig.~\ref{fig:fab}(a). The figure also indicates wavelengths associated with emission lines from the $TX_{0/1}$ levels. We fit Lorentzian lineshapes to the modes to extract the fit parameters.

We observe a structure in these measurements, which is explained by plotting the resonance wavelength of the cavity versus the physical location of the cavity in the array, as in Fig.~\ref{fig:fab}(b). The bullseye cavities on the outside of the array are redshifted, and those on the inside of the array are blueshifted. This indicates that there is room for further improvement of the proximity effect correction in the lithography fabrication step.

We analyse the distribution of cavity parameters in Fig.~\ref{fig:fab}c by looking at the detuning (x-axis) and the linewidth (y-axis). The secondary y-axis on the right indicates the Q factor of the cavities. Each dot is a measured cavity in the array. The dashed line indicates $\Gamma=2|\Delta\lambda|$, the linewidth that optimises $QL$ given a certain detuning. Our performance metric $QL$ is also indicated by the colour scale on the background.

The histograms on the top and on the right show the distributions of cavities and indicate the mean and standard deviations on these metrics. We find averages for the detuning of $\Delta\lambda =(0.0\pm1.1)$~nm and linewidth of $\Gamma=(4.7\pm1.0)$~nm. The average quality factor is $Q=298\pm78$.

\begin{figure}
    \centering
    \includegraphics[width=\linewidth]{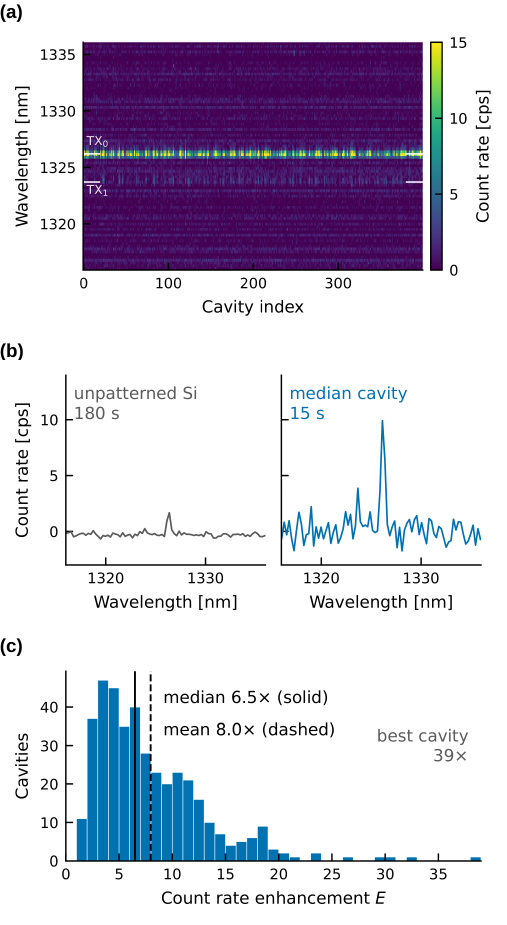}
    \caption{\label{fig:speed}%
    \textbf{T centre signal enhancement and measurement speed-up.} (a) PL spectrum of every cavity (10~$\mu$W, 15~s), T centre ZPL dominates the signal. (b) Comparison between T centre signal in unpatterned silicon (grey, left) and the median-enhancement cavity (blue, right) at the same power, on a common scale. The noise level differs because of a shorter integration time for the cavity spectrum. The count rate occasionally drops below zero because both spectra are background-corrected for spectrometer dark counts. (c) Distribution of signal enhancement $E$ of all 400 cavities vs the unpatterned reference. Because the measurement is background-limited, the required integration time at fixed signal-to-noise ratio (SNR) drops $42\times$. The 400 measurements took 5.5 hours, of which roughly one third was spent on the spectrum acquisition, which is the part that scales with the signal strength. More information on the timing in Appendix Fig.~\ref{fig:app-scans}.}
\end{figure}

\section{Measurement speed-up}
\label{sec:speed}
The degree to which T centre measurements can be sped up can be determined from measuring the increase of signal collected from the ZPL. We therefore measure the PL signal from the same 400 cavities and compare it to the T centre signal collected from unpatterned silicon.

Figure~\ref{fig:speed}(a) shows this measurement at 10~\textmu W of green excitation. A faint $TX_1$ trace is visible, but the signal is dominated by the enhanced $TX_0$ light. Figure~\ref{fig:speed}(b) shows a comparison between the T centre signal from unpatterned Si and that from the cavity with median signal enhancement.

The reference measurements of T centres in unpatterned Si were taken on the same sample, during the same cooldown, on a $3\times3$ grid of nine unpatterned sites. This resulted in a mean reference signal of $1.71\pm0.06$ cps. The separate measurements are shown in Appendix Fig.~\ref{fig:app-scans}(c).

The increased noise level in the cavity spectrum is due to a shorter integration time, which leaves less time for the noise to average out to the baseline. This is a direct example of the advantage that the cavities give. The T centre signal in unpatterned Si would have vanished in the noise with the short integration time used for the cavity measurement. 

The histogram in Fig. \ref{fig:speed}(c) shows the full distribution of cavity performance, indicating the median, mean, and best count rate enhancements. For these measurements, the SNR $\propto E\sqrt{t}$ where $E$ is the cavity count rate enhancement and $t$ is the integration time of the measurement. This means that, for a fixed SNR, $t\propto E^{-2}$.

A median enhancement of $6.5\times$ corresponds to a possible speed-up of $42\times$. With this enhancement, we can decrease measurement time to 15~s per cavity, which is also the integration time used to measure the data in Fig.~\ref{fig:speed}. In our setup, there is a constant 35~s/cavity overhead from checking for latching on the photodetectors, moving between cavities, and position optimisation, which amount to a constant time overhead that does not scale with the enhancement. Detailed timing measurements are included in Appendix Fig.~\ref{fig:app-scans}(b).

This means that, for our setup, the measurement is not limited by the T centre signal anymore. Further speedup can be achieved in several ways. Checking for latching can be done less frequently: checking after every cavity is excessive, it could also be done once every tens of cavity measurements. Stage motion can also be optimised: moving from a raster scan to a snake scan would save a little, but foregoing the position optimisation entirely could save a lot of time. Appendix Fig.~\ref{fig:app-objective} shows that the optimisation shifts the objective position by roughly 1~\textmu m, which is well within the range where cavity light can be collected, but this only works if the exact signal strength per cavity is not of interest.

\section{Statistical analysis of the spread}
\label{sec:spread}
The ZPL signal measured from a cavity is determined by two things: the cavity performance metric $QL$ and the placement of T centres inside the cavity (in number, position, and orientation). The formation of T centres is a probabilistic process that cannot be controlled at the moment, though progress has been made that limits the uncertainty to a pulsed laser spot size \cite{andrini_activation_2024,jhuria_programmable_2024}. We therefore expect the ZPL signal to correlate with $QL$, but not perfectly.

\begin{figure}
    \centering
    \includegraphics[width=\linewidth]{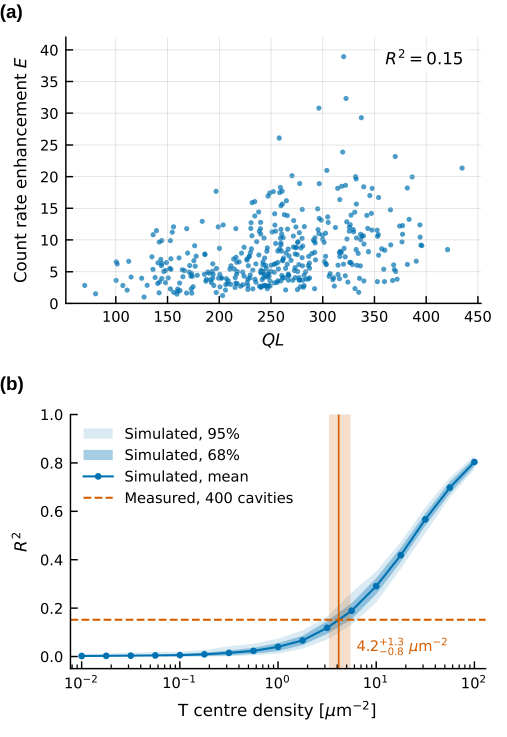}
    \caption{\label{fig:spread}%
    \textbf{Cavity statistical analysis.} (a) Measured T centre signal enhancement $E$ versus $QL$. The correlation is weakly positive ($R^2=0.15$) because T centre location varies randomly. (b) Monte Carlo simulation of the $R^2$ value for varying T centre density. The measured $R^2=0.15$ (orange dashes) is reproduced at a T centre density of $4.2^{+1.3}_{-0.8}$~\textmu m$^{-2}$ (orange line and shading).}
\end{figure}

In this section we use the measured statistics to infer the T centre density in the sample, estimation of which otherwise depends strongly on assumptions about the collection efficiency of the setup used \cite{macquarrie_generating_2021}. We start from the spread of cavities shown in Fig.~\ref{fig:spread}(a), which plots the performance $E$ versus the $QL$ metric. The measurements show a weakly positive correlation of $R^2=0.15$.

We set up a Monte Carlo (MC) simulation to predict the value for $R^2$ as a function of T centre density in the material. Considering the extreme scenarios gives an intuition for why this works. If there are very few T centres in the sample, we expect there to be virtually no correlation, because the stochastic T centre presence/coupling determines the performance almost independently of cavity parameters. In the limit of high T centre density, the distribution of centres will be so homogeneous that we expect the correlation to be almost perfect.

The MC simulation starts from the measured distribution of cavity parameters and then samples T centres randomly throughout the material for a certain density. The signal is calculated by adding the contributions of the individual T centres, based on their location and orientation in the cavity. The value for $R^2$ is calculated from this data. This process is repeated multiple times to estimate a probability distribution for $R^2$. Figure~\ref{fig:spread}(b) shows the result. The graph intersection at $R^2=0.15$ gives a T centre density of $4.2^{+1.3}_{-0.8}$~\textmu m$^{-2}$. We emphasize that this estimated value is independent of the collection efficiency of the system. The Appendix takes a deeper dive into the assumptions underlying the simulation and their justification.

\section{Conclusion and outlook}
We have shown that the need for individual cavity tuning, which has so far limited T centre studies to a handful of emitters, can be removed by designing low-Q cavities that are resonant without individual tuning. We established a design principle of $\Gamma \approx 2\sigma_{\Delta\lambda}$, matching the cavity linewidth to roughly twice the fabrication spread of the cavity wavelength detuning.

We illustrated this by fabricating an array of 400 bullseye cavities with detuning from the T centre of $0.0\pm1.1$~nm, and quality factors of $298\pm78$. These were used to measure 400 spectra of T centre ensembles, showing a median $6.5\times$ signal enhancement, leading to a $42\times$ speedup of the measurement. As a result, we were able to run the measurement in 5.5 hours, of which 1 hour and 40 minutes (less than a third) was actually spent on integration time for the spectra.

The measurement can be sped up further by reducing the overhead. Figure~\ref{fig:fab}(b) also suggests the detuning spread can be reduced by improving the proximity effect correction. In that case, it also becomes possible to start working with cavities with higher quality factors than bullseye cavities can reach. An option could be to look at 2D photonic crystal cavities \cite{saggio_cavity-enhanced_2024}, and deliberately reduce their Q in the design phase.

We used the statistics of the PL signal to determine the T centre density in the material, independent of collection efficiency. The same approach can be extended to PLE measurements to study the linewidths of individual T centres. These measurements provide a route to a statistical understanding of how sample processing affects the formation and properties of T centres. This understanding is crucial for scaling current proof-of-principle experiments up to large numbers of working T centre devices on a single chip.

\section*{Acknowledgements}
We thank Yanik Herrmann and Régis Méjard for support in building the experimental setup, and Nina Codreanu for help with fabricating early versions of the bullseye cavity structures. M.P., E.V., and C.E.-H. acknowledge financial support from the Dutch Research Council (Projects 601.QT.001 and NGF.1623.23.027). R.H. and C.F.P. acknowledge financial support from the Dutch Research Council (NWO) through the Spinoza prize 2019 (project number SPI 63-264) and from the Kavli Foundation through the QM4QT project.

\section*{Author contributions}
C.F.P., M.P., and E.V. designed the sample layouts. M.P. fabricated the samples and calibrated the process. C.F.P., J.M.R., and C.C.P.E.H. set up the T centre creation process and analysed early versions of samples. C.F.P. and B.Z. carried out measurements for the paper. R.H. and C.E.-H. supervised the experiments.

\section*{Data availability statement}
Data will be made available at \cite{Primavera2026Data}.

\bibliography{bibliography}

\clearpage
\appendix
\renewcommand{\thefigure}{S-\arabic{figure}}
\setcounter{figure}{0}

\section{Bullseye cavity simulation}
\label{app-simulation}

\begin{figure}[h]
    \centering
    \includegraphics[width=\linewidth]{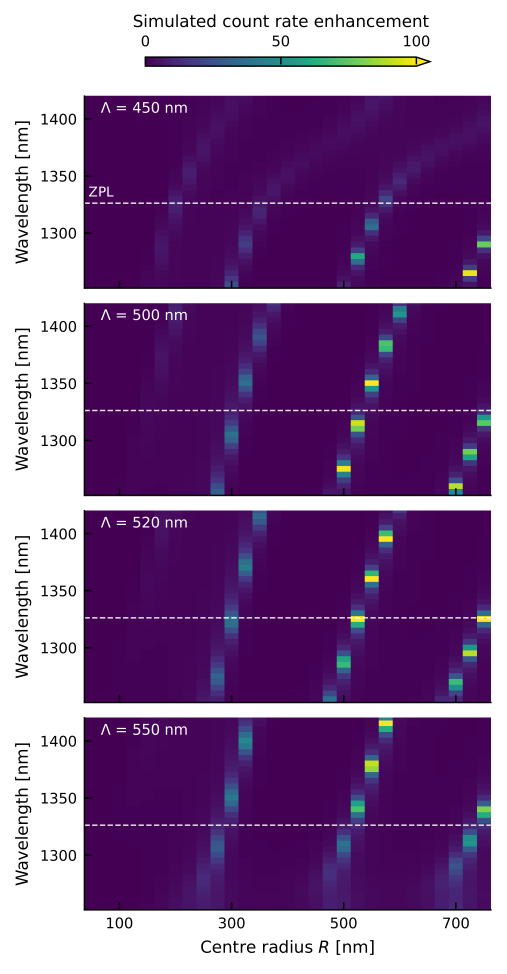}
    \caption{\label{fig:app-periods} \textbf{Simulation checks.} Overview of mode simulation for varying the period $\Lambda$. All bullseye cavities use N=3, $\ff=0.8$.}
\end{figure}

The effect of changing the bullseye cavity grating period $\Lambda$ is illustrated in Fig. \ref{fig:app-periods}. Each period yields a range of wavelengths for which good cavity performance is possible. The radius can be used to select the exact resonance wavelength. Changing the period leads to a shift in this range.

\begin{figure}[h!]
    \centering
    \includegraphics[width=\linewidth]{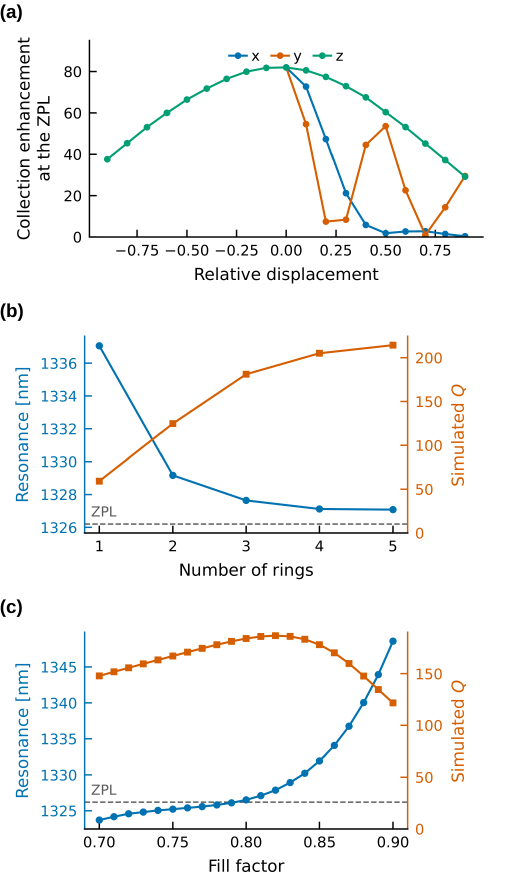}
    \caption{\label{fig:app-simulation} \textbf{Simulation checks.} (a) Simulated collection enhancement at the T-centre ZPL as the dipole is displaced from the cavity centre along $x$, $y$ (in plane) and $z$ (out of plane); displacement in $x$ and $y$ is relative to the central radius $R$, in $z$ relative to the 220~nm device-layer thickness. 0 is defined as the centre of the disc, in the middle of the device layer. $\Lambda=500$~nm, $R=535$~nm, $\ff=0.8$. (b) Fitted resonance wavelength and Q factor as a function of number of rings. When increasing the number of rings, both measures converge to a limit. From 3 rings onwards, there are other uncertainties that are larger than the deviation caused by the limited number of rings. (c) Resonance wavelength and Q factor as a function of fill factor. The value of Q has an optimum around $\ff$=0.82. Q drops faster when accidentally fabricating larger $\ff$. Also, for larger $\ff$, the resonance wavelength becomes more sensitive to statistical variation in $\ff$. Choosing 0.8 is therefore on the safe side.}
\end{figure}
Figure \ref{fig:app-simulation} shows the result of changing various parameters on the results from the simulation. The effects of displacing the dipole from the centre of the bullseye, changing the number of rings, and changing the fill factor are shown. Based on these results, we choose N=3 as the number of rings to use in further simulations, since increasing the number of rings changes the results only slightly, while significantly increasing the computational load. It also motivates the choice for fixing $\ff$=0.8.

\begin{figure}[h]
    \centering
    \includegraphics[width=\linewidth]{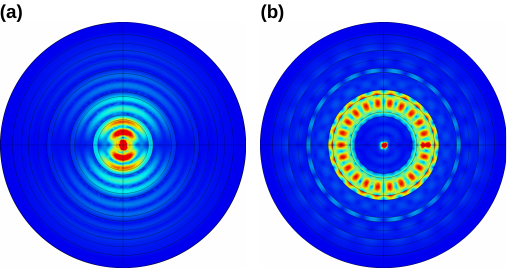}
    \caption{\label{fig:app-modes} \textbf{Mode profiles.} (a) shows the E-field distribution (top view) for the desired modes. (b) shows an example of an unmatched mode, where the excitation dipole has been moved into the first ring.}
\end{figure}

The measured modes that cannot be matched to any of the simulated modes can be recreated by moving the dipole to one of the outer rings. These can be seen as leaky whispering-gallery-modes in the bullseye cavity rings. These don't show up in the simulations because those only take a centred dipole into account. In reality, fluorescence can originate from anywhere in the entire device layer, and some of these modes radiate partially into the collecting objective. Moving the simulated dipole into the rings recreates these modes, as shown in Fig. \ref{fig:app-modes}.

\section{Sample and T centre creation}
All samples in this paper originate from a photonic-grade commercially available CZ grown, P-type, boron doped SOI wafer from Shin-Etsu Handotai Europe. The silicon device layer is 220~nm thick, on top of 3 $\mu m$ of buried oxide.

The sample used for the sweep was implanted with Carbon-12 ions at 38 keV, dose of 7E13 /cm$^2$, at 7 degree tilt angle. It was annealed in argon atmosphere at 1000 degC for 20s. It was then implanted with Hydrogen-1 ions at 9 keV, dose of 7E13 /cm$^2$, at 7 degree tilt angle. The final annealing was done in nitrogen atmosphere at 550 degC for 3 minutes.

The sample containing 400 identical bullseye cavities had the same implantation recipe, but with both doses set to 7E12/cm$^2$, since it was clear in the first sample that there were more T centres than required, even though no exact estimate was known.

Also, the second annealing was executed at 485 degC for 3 minutes in nitrogen atmosphere. The change of temperature results from using a different rapid thermal annealer, which resulted in a different process optimisation result.

The T centre creation recipe was used from \cite{bergeron_silicon-integrated_2020}, and adapted to the machines available to us in the cleanroom.

\begin{figure}[h]
    \centering
    \includegraphics[width=\linewidth]{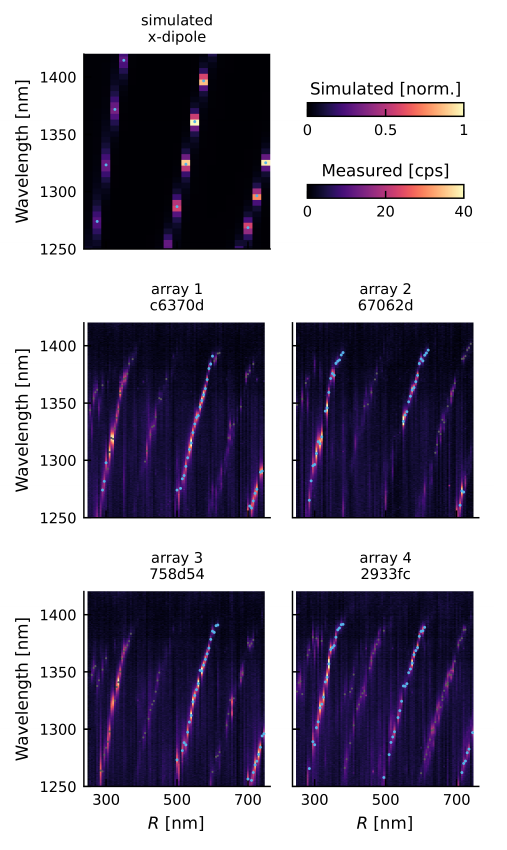}
    \caption{\label{fig:app-sweep-maps} \textbf{All radius-sweep arrays.} Top panel: simulated collected power ($x$-dipole, $\Lambda=520$~nm, normalised). Other panels: measured high-power maps of the four arrays ($R=250$--745~nm in 5~nm steps, $\Lambda=520$~nm, $\ff=0.8$). Dots: fitted resonances.}
\end{figure}

\section{Bullseye cavity fabrication}
Patterns were defined by spinning AR-P 6200.13 resist to a target thickness of $400$~nm, followed by a bake at 150~$^\circ$C for 3~minutes. Electron beam lithography was performed with a base dose of 275~$\mu$C/cm$^2$. Development was carried out in pentylacetate for 1~minute, stopped with a 5~second xylene bath, and rinsed in IPA. Pattern transfer into the silicon substrate was performed by deep reactive ion etching (DRIE) in an Oxford ICP system using a HBr/Ar chemistry with end-point detection. Resist stripping was performed in DMF at 80$^\circ$C for 2~hours, followed by piranha cleaning.

\begin{figure}[h]
    \centering
    \includegraphics[width=\linewidth]{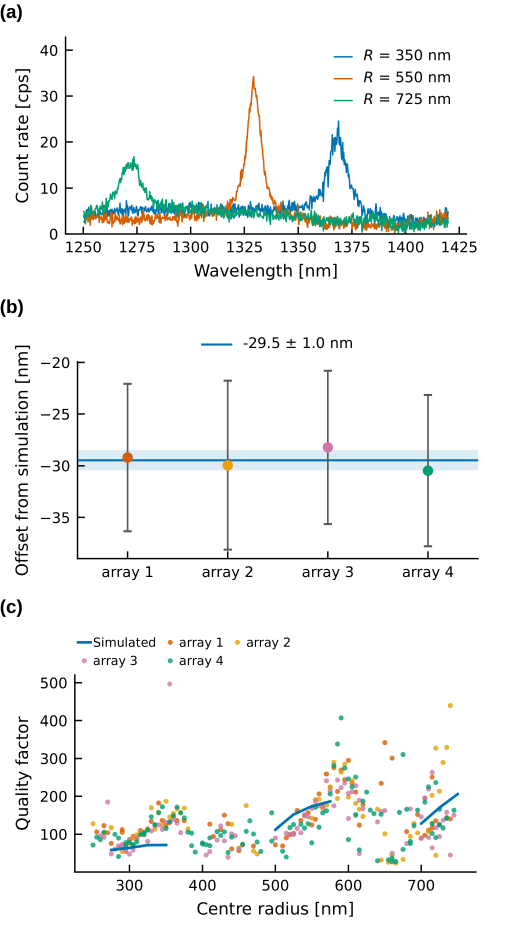}
    \caption{\label{fig:app-sweep-stats} \textbf{Radius sweep analysis.} (a) Example spectra of three cavities of array 3. (b) Offset of the matched resonances from simulation for each array (error bars: r.m.s.\ scatter within the array); line and band: mean and spread over arrays, $-29.5\pm1.0$~nm. (c) Quality factor of every detected mode vs radius, with the simulated $Q$ (line).}
\end{figure}

\begin{figure}
    \centering
    \includegraphics[width=\linewidth]{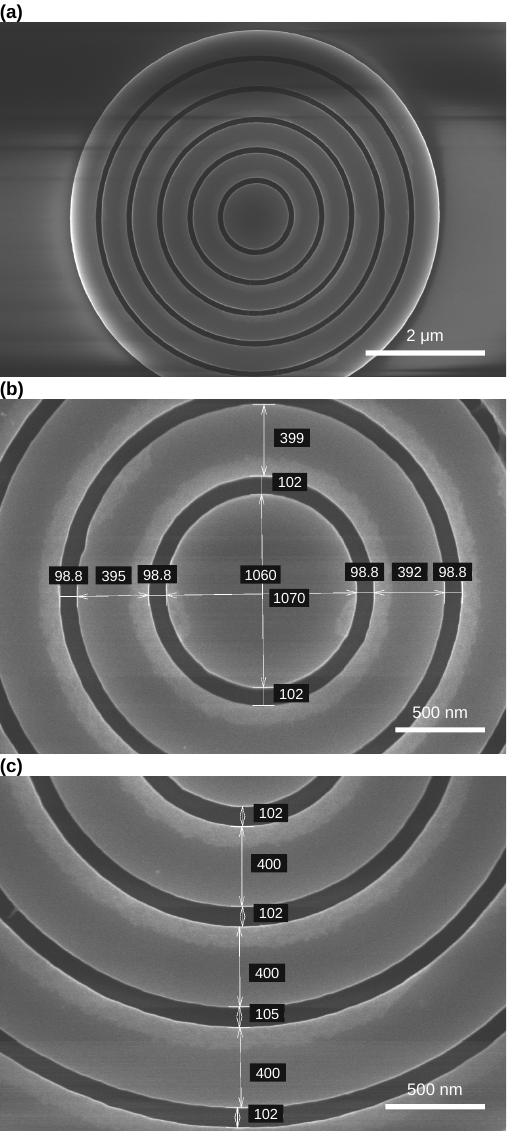}
    \caption{\label{fig:app-device} \textbf{Fabricated device.} (a) SEM of one fabricated cavity from the final array of 400 cavities. (b-c) zoomed in SEM pictures of the final bullseye cavity structure with measurements. Uncertainties are 3 nm because of limited precision in the image analysis. It's clear that the final dimensions for the rings (with $\Lambda=500$ nm and $\ff=0.8$) agree with the design: resp. 400 and 100 nm for the rings and trenches. However, the centre radius comes out to 530-535 nm, exactly like the simulation, but very much different from the 562.5 nm drawn in the design. This confirms that the simulations agree well with reality, and that the offsets are caused during the transfer of the design onto the sample.}
\end{figure}

\section{Optical setup and measurement protocol}
\begin{figure}[h!]
    \centering
    \includegraphics[width=\linewidth]{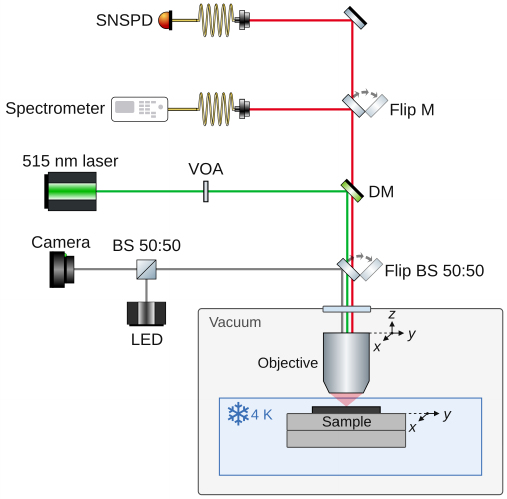}
    \caption{\label{fig:app-setup} \textbf{Experimental setup.} The sample is cooled down to 4 K in a Montana S50 cryostat. Inside is a custom heat shield, with an insert that allows the NA=0.85 objective in the vacuum to get very close to the sample, while still being at room temperature itself. The heat shield insert has a small hole that gives the objective access, while limiting thermal radiation to the cold parts of the cryostat. The objective moves in closed-loop configuration in three dimensions. The sample can be moved over long distances by cryogenic xy-positioners run in open-loop configuration. The imaging system (camera+LED) has optical access to the objective using a flipping pellicle beam splitter, which is flipped in during the optimisation steps, and flipped out during the spectrum acquisition. The excitation and collection path are separated by a DMSP950 dichroic mirror. The schematic shows a long-pass configuration in contrast to the real short-pass configuration. A motorized flip mount from Radiant Dyes switches the collected signal between spectrometer (Princeton Instruments) and SNSPD (Photonspot). The green laser is a fiber-pigtailed Hübner Photonics Cobolt 06-01 at 515 nm.}
\end{figure}

Measurements were carried out in a cryogenic confocal microscopy setup, as shown in Fig. \ref{fig:app-setup}. The measurement takes the locations of the corners of the array as an input, along with the side lengths of the array in number of bullseye cavities. It goes through a sequence of checking for a latch on the SNSPDs (superconducting nanowire single photon detectors), moving the objective to a bullseye, optimising the objective position in x, y, and z directions sequentially, and then taking a spectrum on the bullseye cavity.

A separate background spectrum is recorded once with the laser off, to subtract the dark counts in the data analysis. The background level drifts during the measurement, so the background is multiplied with a different scale factor for each measurement before subtraction, using a least-squares calculation.

The unpatterned reference is measured on the same sample in the same setup, with a very similar measurement. The difference is that only the z-axis of the objective is optimised, and that the integration time is 3 minutes per point, to be able to distinguish the T centre signal from the noise. 9 sites are chosen on the sample, and the average is used as the reference figure of 1.71 cps.

\begin{figure}
    \centering
    \includegraphics[width=\linewidth]{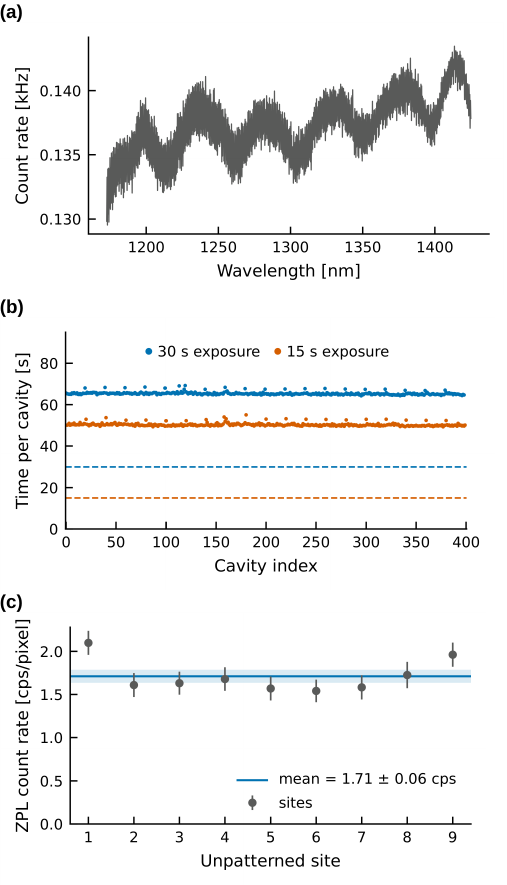}
    \caption{\label{fig:app-scans} \textbf{Experimental details.} (a) Detector background subtracted from the high-power scan. The detector background is scaled with a constant to account for slow drifts in the strength of the background over time. (b) Time per cavity in the high-power (30~s, blue) and low-power (15~s, orange) scans; dashed: exposure time. The difference is a constant $\approx35$~s of stage motion and position optimisation. (c) ZPL count rate at the nine unpatterned reference sites; line and band: mean $1.71\pm0.06$~cps, where $0.06$~cps is standard error.}
\end{figure}

\begin{figure}
    \centering
    \includegraphics[width=\linewidth]{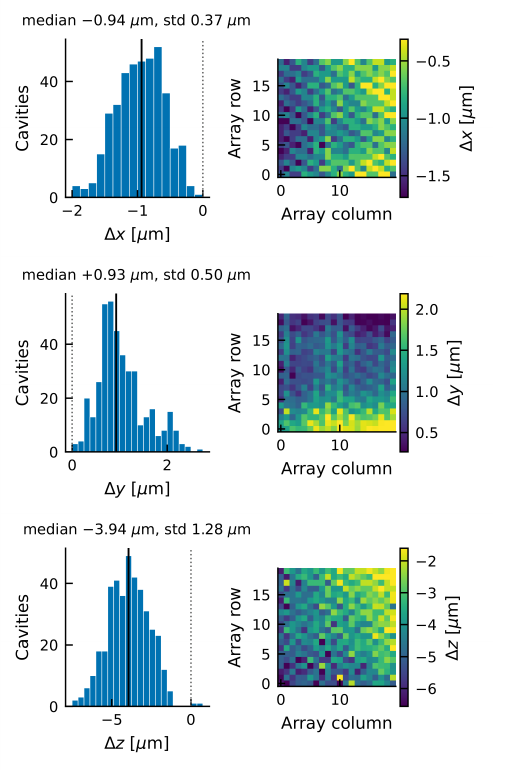}
    \caption{\label{fig:app-objective}\textbf{Optimisation analysis.} Panels show the differences between optimised position and original position, for each axis separately, for the low-power measurement of T centre signal. The histograms show a systematic offset in each of the axes. This is because the positions of the array were determined before the high-power measurement, and not changed afterwards. For running the entire measurement without optimisation in between, it's possible to redefine the positions of the corners, leading to a remaining position deviation with standard deviation $<$ 1 \textmu m for x and y, and $<$ 2 \textmu m in z, which is safely inside the region where signal is collected. The 2D scans, especially for x and y also show an apparent pattern on top of the gradient. This turns out to be because the array positions are rounded to entire micrometers. This is unnecessary and was later removed.}
\end{figure}

Specifics of the protocol are shown in Fig. \ref{fig:app-scans}. Results from the optimisation steps are shown in Fig. \ref{fig:app-objective}.

\section{Data analysis}
Our data analysis builds on internal software infrastructure for analysing and visualising measurement results. When preparing this publication, we used Anthropic's AI tools (specifically Claude Opus 5 and Opus 5.5) to turn this analysis code into a standalone package. All code generated was checked and edited by the authors, who take full responsibility for it. The code that reproduces the figures will be published together with the raw data at \cite{Primavera2026Data}.

\section{Monte Carlo simulation}
The Monte Carlo simulation starts from the distribution of the 400 measured cavity linewidths and detunings. It then stochastically generates T centres based on an assumed density. It calculates the coupling between the T centres and the cavity mode based on their location and orientation. From this, it deduces the expected total T centre signal. After having done this for each cavity, it can calculate a correlation value $R^2$. It does this many times to find the distribution of $R^2$. By intersecting this graph with the measured value of $R^2$, we estimate the T centre density in our measured sample.

\begin{figure}
    \centering
    \includegraphics[width=\linewidth]{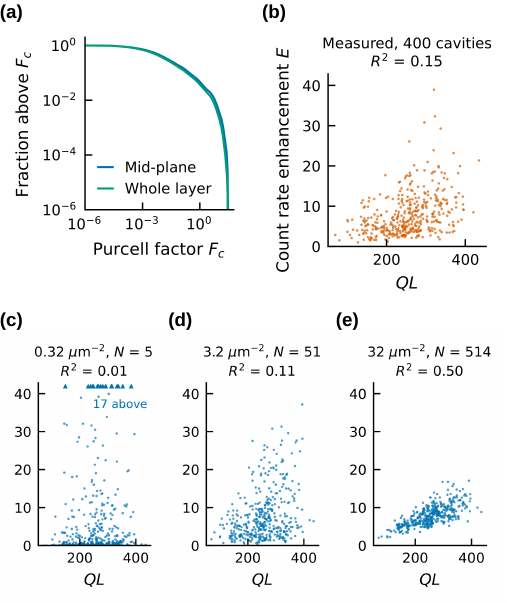}
    \caption{\label{fig:app-montecarlo} \textbf{Monte Carlo simulation.} (a) Distribution of the on-resonance Purcell factor $F_c$ over random emitter positions and dipole orientations in the device layer, at $Q=298$. (b) Measured correlation between $QL$ and $E$. (c-e) Simulated correlations between $QL$ and $E$ for three different densities. For low density, there is barely any correlation, and for higher densities, the correlation becomes very strong.}
\end{figure}

\begin{figure}
    \centering
    \includegraphics[width=\linewidth]{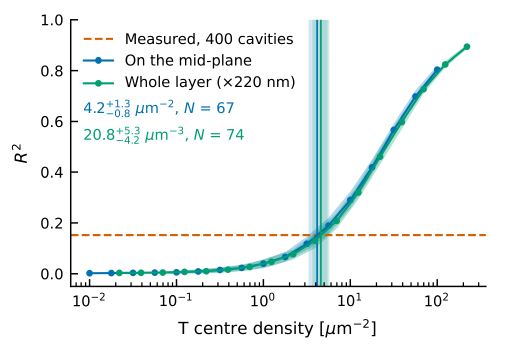}
    \caption{\label{fig:app-placement} \textbf{Monte Carlo z-axis placement.} For the Monte Carlo simulation, we considered two extreme cases of the z-distribution of emitters: on one end is the assumption of a homogeneous distribution (in green), on the other end the assumption that every emitter lies exactly in the middle of the device layer (blue). The real distribution isn't known exactly, but running the simulation for both assumptions shows that the results for the estimated T centre density don't differ from each other significantly.}
\end{figure}

One of the assumptions that goes into the simulation is the distribution of emitters in the z-direction. The ion implantation is aimed at the middle of the device layer, a depth of 110 nm, with a simulated straggle of 40 nm. However, it's unknown where the emitters actually form. The simulation allows for two options: everything exactly in the middle of the device layer, or a completely homogeneous distribution throughout the entire device layer. It turns out that the results agree with each other within the error margin. These results are shown in Fig. \ref{fig:app-placement}. The simulation allows for multiple choices of how the dipole orientations of the emitters are chosen. This also doesn't significantly change the results.

\begin{figure}
    \centering
    \includegraphics[width=\linewidth]{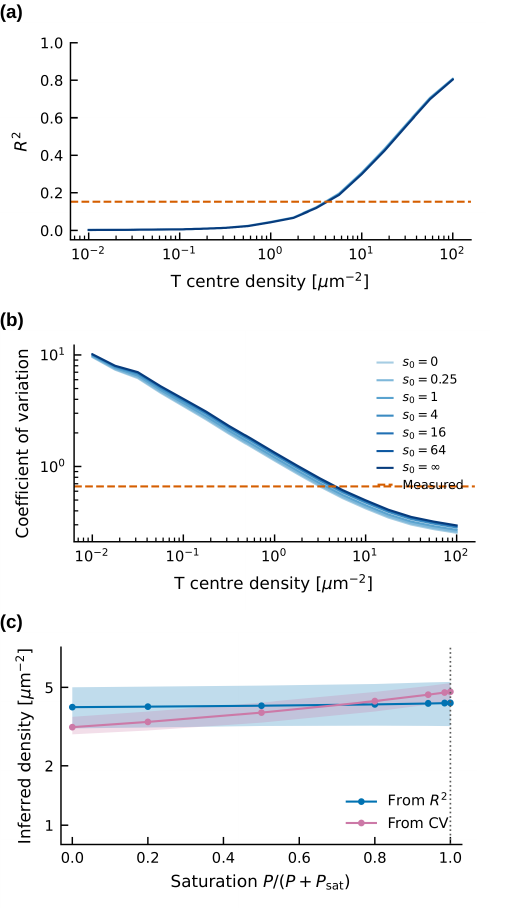}
    \caption{\label{fig:app-saturation} \textbf{Monte Carlo assumption variation.} It is possible to choose a different measure to relate the density to the measured statistics, for example the count rate variability CV. (a) and (b) show a comparison between the two measures. Also, the level of saturation of the emitters is varied, indicated by the graphs in different shades of blue. In the case of $R^2$, this doesn't impact the results significantly, so the lines overlap. (c) shows the inferred density for both measures as a function of $s_0=P/P_{sat}$. All results agree with each other within error. Only the results using CV seem to be sensitive to the saturation level.}
\end{figure}

Similarly, the simulation shows that choosing a different measure that varies with density doesn't change the result. Neither does the degree of saturation of the emitters. Figure \ref{fig:app-saturation} shows this.

\clearpage

\end{document}